\documentclass[conference]{IEEEtran}
\IEEEoverridecommandlockouts
\usepackage{bm}
\usepackage{cite}
\usepackage{amsmath,amssymb,amsfonts}
\usepackage{algorithm}
\usepackage{algpseudocode}

\newcommand{\customsubsection}[1]{%
  \begingroup
  \centering
  \small\scshape #1\par
  \endgroup
  \vspace{1.5ex}
}
\usepackage{newtxtext}

\usepackage{graphicx}
\usepackage{textcomp}
\usepackage{xcolor}
\def\BibTeX{{\rm B\kern-.05em{\sc i\kern-.025em b}\kern-.08em
    T\kern-.1667em\lower.7ex\hbox{E}\kern-.125emX}}
\begin{document}

\title{End-Edge Model Collaboration: Bandwidth Allocation for Data Upload and Model Transmission}

\author{
Dailin Yang\IEEEauthorrefmark{1},
Shuhang Zhang\IEEEauthorrefmark{2},
Hongliang Zhang\IEEEauthorrefmark{2}, and
Lingyang Song\IEEEauthorrefmark{2}\IEEEauthorrefmark{3}\IEEEauthorrefmark{4}
\\[0.15in]
\IEEEauthorblockA{
\IEEEauthorrefmark{1}School of Electronics Engineering and Computer Science, Peking University, Beijing, China, 100871 \\
\IEEEauthorrefmark{2}School of Electronics, Peking University, Beijing, China, 100871 \\
\IEEEauthorrefmark{3}School of Electronic and Computer Engineering, Peking University Shenzhen Graduate School, Shenzhen, China, 518055 \\
\IEEEauthorrefmark{4}Hunan Institute of Advanced Sensing and Information Technology, Xiangtan University, Xiangtan, China, 411105
}
}

\maketitle

\begin{abstract}
The widespread adoption of large artificial intelligence (AI) models has enabled numerous applications of the Internet of Things (IoT). However, large AI models require substantial computational and memory resources, which exceed the capabilities of resource-constrained IoT devices. End-edge collaboration paradigm is developed to address this issue, where a small model on the end device performs inference tasks, while a large model on the edge server assists with model updates. To improve the accuracy of the inference tasks, the data generated on the end devices will be periodically uploaded to edge server to update model, and a distilled model of the updated one will be transmitted back to the end device. Subjected to the limited bandwidth for the communication link between the end device and the edge server, it is important to investigate whether the system should allocate more bandwidth to data upload or to model transmission. In this paper, we characterize the impact of data upload and model transmission on inference accuracy. Subsequently, we formulate a bandwidth allocation problem. By solving this problem, we derive an efficient optimization framework for the end-edge collaboration system. The simulation results demonstrate our framework significantly enhances mean average precision (mAP) under various bandwidths and datasizes.  
\end{abstract}


\section{Introduction}
\begin{figure*}[t]
  \centering
  \includegraphics[width=0.7\textwidth,height=0.3\textwidth,keepaspectratio=false]{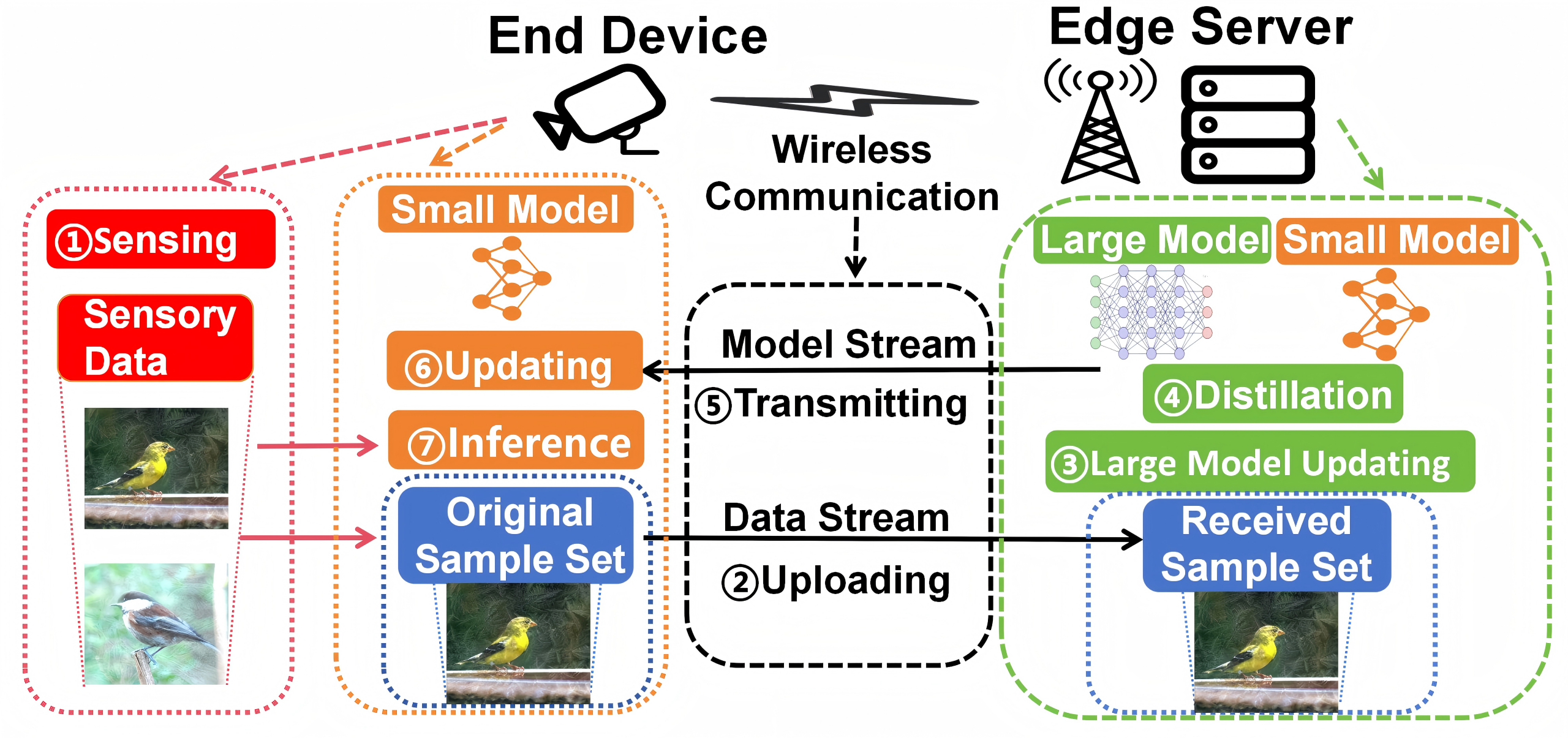}
  \caption{System model for end-edge collaboration}
  \label{fig:framework}
\end{figure*}

The rapid development of the large artificial intelligence (AI) models enables both high accuracy and fast inference in practical tasks, which are crucial for real-time inference in applications of the Internet of Things (IoT)\cite{Shao2023}. Deploying AI models on end devices, where data is generated, further mitigates cloud-related latency. However, the computational and memory constraints of these devices restrict the deployment of large models. To address this challenge, a model collaboration paradigm has been proposed . In such a paradigm, an end device runs a small model for real-time inference, and an edge server hosts a large model. The end device uploads the sensory data to the edge server. The edge server utilizes the received data to perform model updates by knowledge distillation and transmits the distilled model to the end device. The transmission processes are as follows:
\begin{itemize}
    \item \textbf{Data Upload}: Adjusts the volume and quality of upload sensory data to balance model update effects and transmission overhead.
    \item \textbf{Model Transmission}: Adjusts the number of transmitted model parameters to maximize inference performance under bandwidth limits.
\end{itemize}

In literature, several studies have explored the model collaboration paradigm. Authors in \cite{Zhang2025} introduced a new integrated air-ground edge-cloud model evolution framework. In \cite{Yu2024}, an innovative end-edge collaborative model framework was proposed that empowers both end nodes and edge servers
to jointly engage in real-time model computation. The work in \cite{Luo2020} developed an optimal uplink resource allocation strategy that assigns bandwidth and computing capacity to devices. Authors in \cite{Su2025} designed an adaptive uplink resource allocation mechanism to minimize transmission delay and energy usage. Although prior works \cite{Luo2020,Su2025} concentrated on resource allocation for data upload, they overlooked the crucial interplay between data upload and model transmission, which is essential for maintaining overall system performance. 

The real-time collaboration between the end device and the edge server is carried out under limited bandwidth\cite{Deng2022}. Therefore, the system should optimize its bandwidth allocation for data upload and model transmission. In this paper, we aim to maximize the mean average precision (mAP) of the end-edge collaboration system by jointly optimizing the data upload and model transmission under specific communication constraints.

However, such an optimization faces a new challenge, i.e., the impact of the number of transmitted model parameters on mAP remains unclear. To address this challenge, we conducted a series of simulations to evaluate the mAP with various numbers of model parameters transmitted to the end device. Based on the results, we fit a function characterizing the impact of the number of transmitted model parameters on mAP. Subsequently, we formulate a bandwidth allocation optimization problem that maximizes the mAP of the end-edge collaboration system.
We derive an efficient communication resource allocation algorithm to solve this problem and further investigate the trade-off between data upload and model transmission under various bandwidths. 

The rest of this paper is organized as follows: Section II details the system architecture and mathematical models. Section III presents the optimization problem and its decomposition. Section IV derives the solution to the optimization problem and analyzes it. Section V validates the performance through comparative simulations and investigates the trade-off between uplink and downlink data transmission, followed by conclusions in Section VI.

\section{System Model}
\subsection{System Overview}
The end-edge collaboration system comprises \textbf{an end device} and \textbf{an edge server}, as illustrated in Fig.~\ref{fig:framework}. The end device collects sensory data and performs inference tasks using a small AI model. However, due to limited computational and memory resources, it is unable to update the model autonomously. To address this challenge, an edge server assists with model updates by training a large model based on the uploaded data and transmitting distilled knowledge back to the end device. 

The system operates as follows: 
\begin{enumerate}
    \item The end device captures sensory data.
    \item A portion of the collected data is uploaded to the edge server for model update.
    \item The edge server updates its large model based on the uploaded data.
    \item The edge server generates model update data through knowledge distillation.
    \item The end device utilizes the update data and performs inference.
\end{enumerate}

Due to limited communication resource, only a proportion of sensory data is uploaded, forming the uplink \textit{data stream}. And only a proportion of the model updates can be transmitted, forming the downlink \textit{model stream}. The trade-off between these two streams significantly impacts the inference accuracy of the edge device, motivating an optimization problem for communication resource allocation.

\subsection{Mathematical Modeling}

\subsubsection{Communication Constraints}
The system utilizes a dedicated channel with fixed bandwidth \( B \). Transmission follows a time-division mechanism. The total available communication time is \( T_{\text{total}} \), allocated between uplink (\( T_u \)) and downlink (\( T_d \)). The uplink spectrum efficiency is denoted as \( S_u \), and the downlink spectrum efficiency is denoted as \( S_d \). In this paper, \( B \), \( S_u \), \(S_d\) are assumed to be constant. Therefore, the system can upload \(BS_u\) bits or download \(BS_d\) bits per second.

The uplink transmission rate $R_V$ and the downlink transmission rate $R_M$ are limited as 
\begin{equation}
R_VT_{total} \leq B S_u T_u.
\end{equation}
\begin{equation}
R_MT_{total} \leq B S_dT_d.
\end{equation}

\subsubsection{Computing Constraints}
The end device captures image data at a frame rate \( N \). Each uploaded frame consists of \( F \) parameters, and the quantization level is denoted by \( q \), which is selected from set \(\bm{\Omega}\). Denote the proportion of sensory data uploaded as \( \rho \). The uplink transmission rate is determined by the volume of uploaded data:
\begin{equation}
R_V = N \rho  F  q.
\end{equation}

Denote the number of model parameters transmitted as $M$. The downlink transmission rate is determined by the number of updated model parameters transmitted to the end device:
\begin{equation}
R_M = \frac{M  b}{T_{\text{total}}}.
\end{equation}
where $b$ is a given positive integer representing the number of
bits allocated to each model parameter in the downlink transmission. Let $M_{\text{max}}$ denote the total number of model parameters, which is the upper bound of $M$.

\subsubsection{Inference Performance Dependency}
The inference performance of the end model is evaluated using mAP, which depends on both the volume of the upload data and the number of the transmitted model parameters. Although a mathematical formulation describing the relationship between variables \( \rho \), \( M \), \( q \), and the \( mAP \) of the end model is not available, we can approximate the dependency in two steps based on existing studies in \cite{Chen2022,Aguilar2020}. 

First, the function \( g(\rho, q) \) captures the dependency of update performance of the edge model on the uplink variables \( \rho \) and \( q \). Denote the update performance as $mAP^*$, then:
\begin{equation}
mAP^* = g(\rho, q).
\end{equation}

Second, the $mAP$ of the end model can be expressed as a function of the effectiveness of on-device model updates and the number of updated model parameters transmitted in the downlink:
\begin{equation}
mAP = f(M, mAP^*).
\end{equation}

Both function \( g(\cdot) \) and function \( f(\cdot) \) vary based on the model architecture and dataset properties and are fitted using experimental data. Given the functions \( g(\cdot) \) and \( f(\cdot) \), the joint impact of uplink and downlink data on the edge model update can be expressed accordingly. 

\section{Problem Formulation and Decomposition}
\subsection{Problem Formulation}

To maximize the objective variable \( mAP \), we aim to optimize the allocation of uplink and downlink communication resource. Specifically, we optimize the quantization parameter \( q \) for the uploaded sensory data, the scaling factor \( \rho \), and the number of parameters \( M \) to be updated in the small model. The optimization problem is formulated as follows:

\begin{subequations}
\label{eq:simplified_objective}
\begin{alignat}{2}
\mathop{\max}\limits_{ M, \rho, q} \quad & mAP = f(M, mAP^*), \label{eq:simplified_main} \\
\text{s.t.} \quad 
& mAP^* = g(\rho, q), \quad 0 \leq \rho \leq 1, \ q \in \bm{\Omega} \label{eq:simplified_constraint_a} \\
& T_{d} + T_{u} \leq T_{\text{total}}, && \label{eq:simplified_constraint_d} \\
& B \cdot S_u \cdot \frac{T_{u}}{T_{\text{total}}} \geq N \rho F q, && \label{eq:simplified_constraint_e} \\
& B \cdot S_d \cdot \frac{T_{d}}{T_{\text{total}}} \geq \frac{M b}{T_{\text{total}}}, && \label{eq:simplified_constraint_f} \\
& 0 \leq M \leq M_{\text{max}}. && \label{eq:simplified_constraint_g}
\end{alignat}
\end{subequations}

The objective function in \eqref{eq:simplified_main} represents the maximization of the $mAP$ for the end-edge collaboration system, which depends on the variables \( M \) and \( mAP^* \). Constraint \eqref{eq:simplified_constraint_a} captures the dependency of the edge model’s update performance on variables \( \rho \) and \( q \). The upload proportion \( \rho \) lies within the range \([0, 1]\) and the quantization parameter \( q \) is selected from the discrete set \( \bm{\Omega} \). Constraint \eqref{eq:simplified_constraint_d} limits the total total available uplink and downlink time. Constraints \eqref{eq:simplified_constraint_e} and \eqref{eq:simplified_constraint_f} define the data size requirements for the uplink data stream and downlink model stream, respectively. Constraint \eqref{eq:simplified_constraint_g} limits the number of parameters \( M \) that are updated in the small model.

\subsection{Problem Decomposition}

Problem \eqref{eq:simplified_objective} presents a major challenge for direct solution: both \(f(\cdot)\) and \(g(\cdot)\) are unknown bivariate functions, which complicates the optimization process. To address this challenge, we identify their key properties to decompose problem \eqref{eq:simplified_objective}, rendering its solution both feasible and practical.

\textbf{Remark 1:} For all \(M \in (0, M_{\text{max}}]\), \(mAP\) strictly increases with respect to \(M\) when \(mAP^*\) is fixed. Accordingly, the fitted function \(h(M, mAP^*)\) satisfies:
\begin{equation}
\frac{\partial f(M, mAP^*)}{\partial M} > 0, \quad 0 < M \leq M_{\text{max}}, \label{eq:remark1}
\end{equation}

Remark 1 characterizes the property of $f(\cdot)$ based on the experimental results in \cite{Chen2022}. Let \(g_j(\rho)\) denote the function of \(\rho\) for \(mAP^*\), given a quantization parameter \(q_j\), where \(j \in \{1,2,\dots,|\bm{\Omega}|\}\). We then have \(g_j(\rho) = g(\rho, q)\bigl|_{q=q_j}\).

\textbf{Remark 2:} For all \(j \in \{1,2,\dots,|\bm{\Omega}|\}\), \(g_j(\rho)\) is a concave function.

Remark 2 characterizes the behavior of \(g(\rho, q)\) based on our experiments and the results in \cite{Aguilar2020}. Define the upper boundary function of problem \eqref{eq:simplified_objective} as \(L_M(M)\), which represents the maximum value of \(mAP^*\) for each value of \(M\) within the feasible region of the bivariate function \(f(M, mAP^*)\), subject to constraints \eqref{eq:simplified_constraint_a}--\eqref{eq:simplified_constraint_g}. It traces the upper limit of the feasible region in the \(M\)--\(mAP^*\) plane. Formally,
\begin{equation}
\begin{aligned}
L_M(M) = \max \{ mAP^* \,\mid\, (\rho, q) \;\text{satisfies}\\
\text{constraints \eqref{eq:simplified_constraint_a}--\eqref{eq:simplified_constraint_g}}\}.
\end{aligned}
\end{equation}

Since \( mAP^* \) is a bivariate function, the shape of the feasible region \( C \) in problem \eqref{eq:simplified_objective} remains unknown. Consequently, the function \( L_M(M) \) remains unknown. To address this issue, we further analyze the function \( L_M(M) \) to lay the foundation for solving problem \eqref{eq:simplified_objective}. We introduce Theorem 1 to characterize the properties of \( L_M(M) \).

\textbf{Theorem 1:} The function \( L_M(M) \) is a continuous, monotonically non-increasing, piecewise function defined on \( [0, \min(M_{\text{max}}, \frac{T_{\text{total}} \cdot B \cdot S_d}{b})] \), where each segment is either constant or concave.

\textit{proof:} See Appendix A.

Leveraging the properties of the function \( L_M(M) \) and Remark 1, we introduce the following theorem to refine the solution space for problem \eqref{eq:simplified_objective}.

\textbf{Theorem 2:} The optimal solution \( (M_{\text{opt}}, mAP^*_{\text{opt}}) \) for problem \eqref{eq:simplified_objective} satisfies \( mAP^*_{\text{opt}} = L_M(M_{\text{opt}}) \).

\textit{proof:} See Appendix B.

Using Theorem 1 and Theorem 2, we decompose problem \eqref{eq:simplified_objective} into two subproblems. The first subproblem is formulated as follows:
\begin{subequations}
\label{eq:subproblem1}
\begin{alignat}{2}
\mathop{\max}\limits_{\rho, q} \quad & g(\rho, q), \label{eq:subproblem1_main} \\
\text{s.t.} \quad 
& 0 \leq \rho \leq 1, \quad q \in \bm{\Omega} \label{eq:subproblem1_b} \\
& T_{d} + T_{u} \leq T_{\text{total}}, && \label{eq:subproblem1_c} \\
& B \cdot S_u \cdot \frac{T_{u}}{T_{\text{total}}} \geq N \rho F q, && \label{eq:subproblem1_d} \\
& B \cdot S_d \cdot \frac{T_{d}}{T_{\text{total}}} \geq \frac{M b}{T_{\text{total}}}, && \label{eq:subproblem1_e} \\
& 0 \leq M \leq M_{\text{max}}. && \label{eq:subproblem1_f}
\end{alignat}
\end{subequations}

Based on the obtained $L_M(M)$, the second subproblem is formulated as:
\begin{subequations}
\label{eq:subproblem2}
\begin{alignat}{2}
\mathop{\max}\limits_{M}\quad  &\ mAP = f(M, L_M(M)), \label{eq:subproblem2_main} \\
\text{s.t.} \quad & 0 \leq M \leq \min\left(M_{\text{max}}, \frac{T_{\text{total}} \cdot B \cdot S_d}{b}\right).  \label{eq:subproblem2_b}
\end{alignat}
\end{subequations}

\section{Communication Resource Allocation Algorithm}
Based on Remark 2, for any $q_j \in \bm{\Omega}$, the subproblem \eqref{eq:subproblem1} reduces to a standard concave optimization problem. Similar to $L_{M}(M)$, we define the function $L_{M,j}(M)$ as the upper limits of the feasible region in the $M–mAP^*$ plane given $q=q_j$.

Since $\bm{\Omega}$ is a finite discrete set, we enumerate each value of $\hat{q}_j$ and solve the concave optimization problem in subproblem \eqref{eq:subproblem1} for each $j$. This yields the function $L_{M,j}(M)$, along with the corresponding optimal variables $\rho_{\text{opt},j}(M)$, $T_{d,\text{opt},j}(M)$, and $T_{u,\text{opt},j}(M)$. Next, we select the combination of variables that maximizes $L_{M,j}(M)$ across all $j$, thereby determining $L_M(M)$, $\rho_{\text{opt}}(M)$, and $q_{\text{opt}}(M)$ for the given value of $M$.

With given $L_M(M)$, subproblem \eqref{eq:subproblem2} reduces to finding the maximum of a single-variable function defined on a closed interval. Since $L_M(M)$ is piecewise differentiable, we can efficiently solve this subproblem to obtain the optimal $mAP$, denoted as $mAP_{\text{opt}}$ and the corresponding optimal values $M_{\text{opt}}$, $\rho_{\text{opt}}$, and $q_{\text{opt}}$. Finally, the optimal values of $T_{u,\text{opt}}$ and $T_{d,\text{opt}}$ are determined using the conditions $B \cdot S_d \cdot T_{d,\text{opt}} = M_{\text{opt}} \cdot b$ and $T_{d,\text{opt}} + T_{u,\text{opt}} = T_{\text{total}}$.

Thus, we summarize the solution to problem \eqref{eq:simplified_objective} in Algorithm 1.

\begin{algorithm}
\caption{Communication Resource Allocation Algorithm}
\label{alg:map_optimization}

\begin{algorithmic}[1]
\Require Parameters \( B \), \( S_u \), \( S_d \), \( N \), \(F\), \( T_{\text{total}} \), \( M_{\text{max}} \), set \( \bm{\Omega} \), functions \( g_j(\cdot) \) for \( j \in \{1,2,\dots,|\bm{\Omega}|\} \), function \( f(\cdot) \)
\Ensure Optimal parameters \( M_{\text{opt}} \), \( q_{\text{opt}} \), \( \rho_{\text{opt}} \), \( T_{u,\text{opt}} \), \( T_{d,\text{opt}} \)

\State Initialize \( mAP_{\text{opt}} \gets -\infty \)
\ForAll{quantization parameters \( q_j \in \bm{\Omega} \)}
  \State Solve subproblem \eqref{eq:subproblem1} to obtain \( L_{M,j}(M) \), \( \rho_{\text{opt},j}(M) \), \( T_{d,\text{opt},j}(M) \), \( T_{u,\text{opt},j}(M) \)
\EndFor
\State Set \( L_M(M) \gets \max_j L_{M,j}(M) \), record \( \rho_{\text{opt}}(M) \), \( q_{\text{opt}}(M) \)
\State Solve subproblem \eqref{eq:subproblem2} to obtain \( M_{\text{opt}} \)
\State Compute \( T_{d,\text{opt}} \gets \frac{M_{\text{opt}} \cdot b}{B \cdot S_d} \)
\State Compute \( T_{u,\text{opt}} \gets T_{\text{total}} - T_{d,\text{opt}} \)
\State \Return \( M_{\text{opt}} \), \( q_{\text{opt}} \), \( \rho_{\text{opt}} \), \( T_{u,\text{opt}} \), \( T_{d,\text{opt}} \)
\end{algorithmic}
\end{algorithm}

\section{Experiment Results} 
\subsection{Experimental Setup}
We evaluate the proposed optimization framework using a vision-based classification task on the ImageNet-C dataset. The end device deploys a ResNet-18 model, while the edge server maintains both a ResNet-101 model and a ResNet-18 model. We compare the performance of our optimization framework with three  frameworks: a distributed computing framework (baseline), a Cloud-Edge elastic model adaptation framework (CEMA) and exhausted search for the end-edge collaboration system.

\begin{enumerate}
    \item \textbf{Distributed computing framework:} In this framework, the small model deployed on the end device performs the inference tasks without communicating with the edge server for model updates.
    \item \textbf{CEMA:} In this framework, the end-edge collaboration system is employed. The communication and model updates are facilitated by the method introduced in \cite{Chen2024}. 
    \item \textbf{Exhaustive search:} In this framework, the end-edge collaboration system is employed. The allocation of communication resource, uplink data quantization optimizations, uplink data ratio optimizations, and end model update design are determined by enumerating over $10^9$ candidate variable combinations for mAP maximization. 
\end{enumerate}

\subsection{Performance Analysis}

\begin{figure}[t]
    \centering
    \includegraphics[width=0.36\textwidth,height=0.23\textwidth]{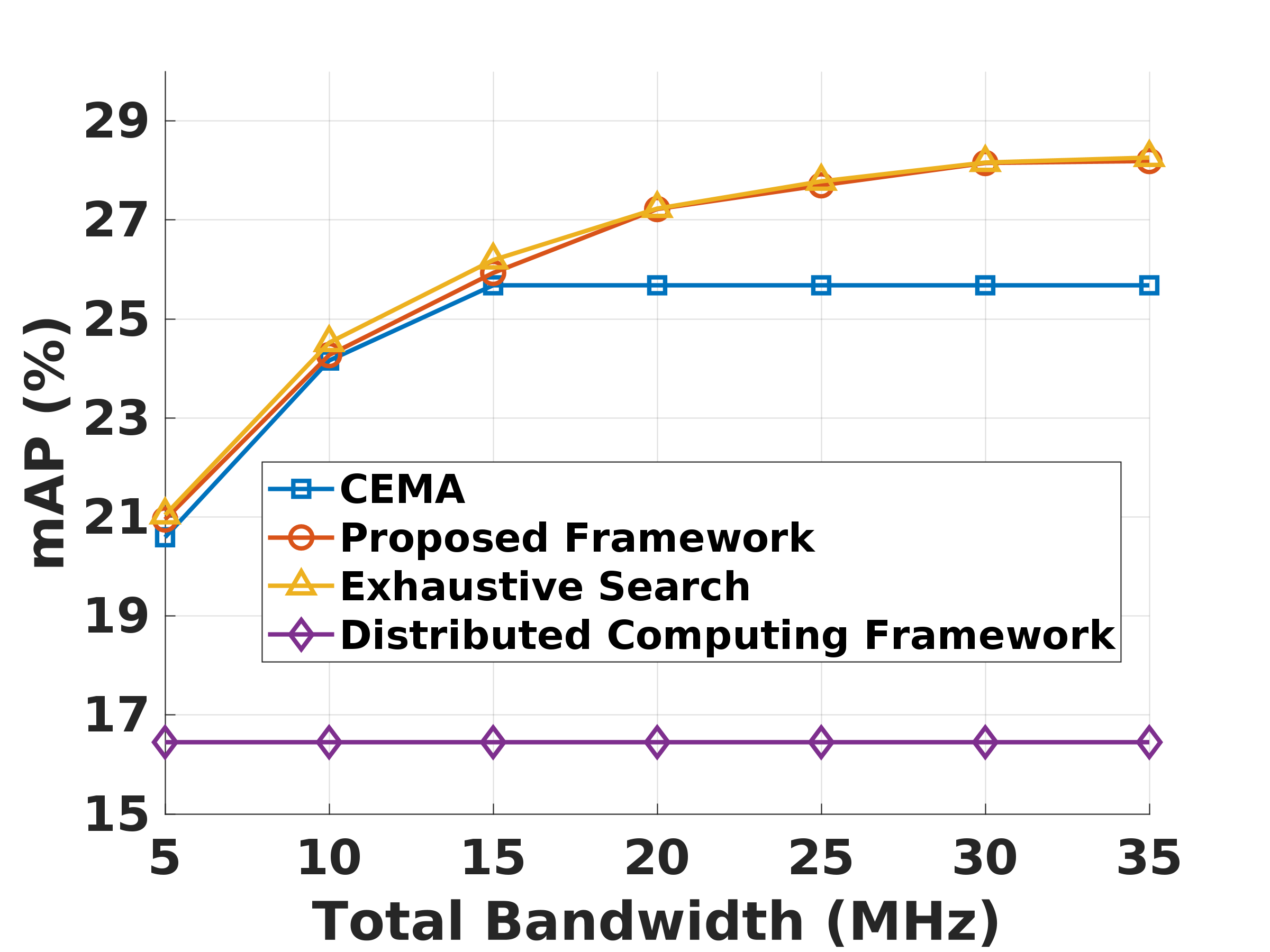}
    \caption{$mAP$ versus total bandwidth $B$ for different methods.}
    \label{fig:bandwidth_map}
\end{figure}

\begin{figure}[t]
    \centering
    \includegraphics[width=0.36\textwidth,height=0.23\textwidth]{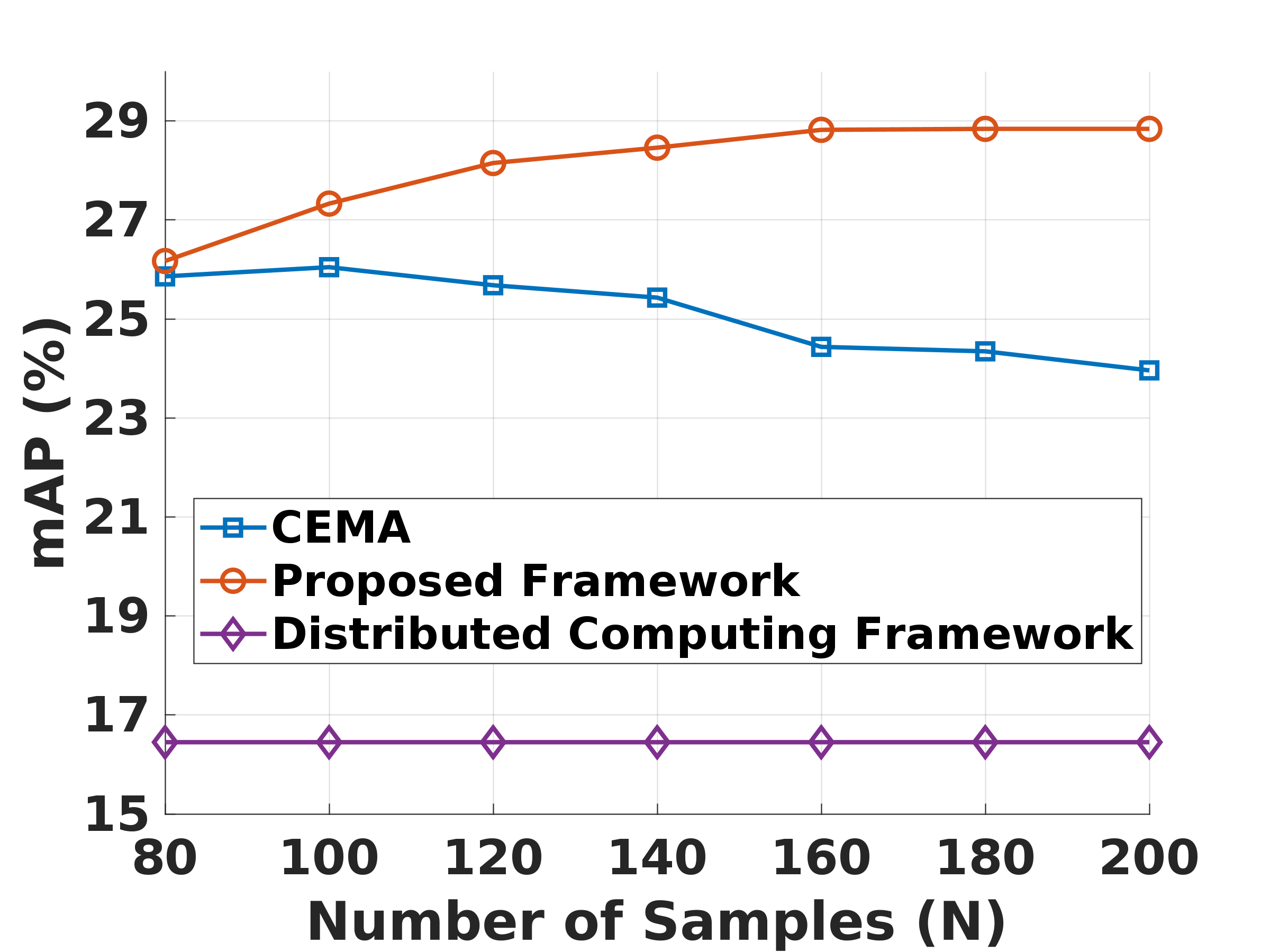}
    \caption{$mAP$ versus frame rate $N$ for different methods.}
    \label{fig:N_map}
\end{figure}

\begin{figure}[t]
    \centering
    \includegraphics[width=0.38\textwidth,height=0.24\textwidth]{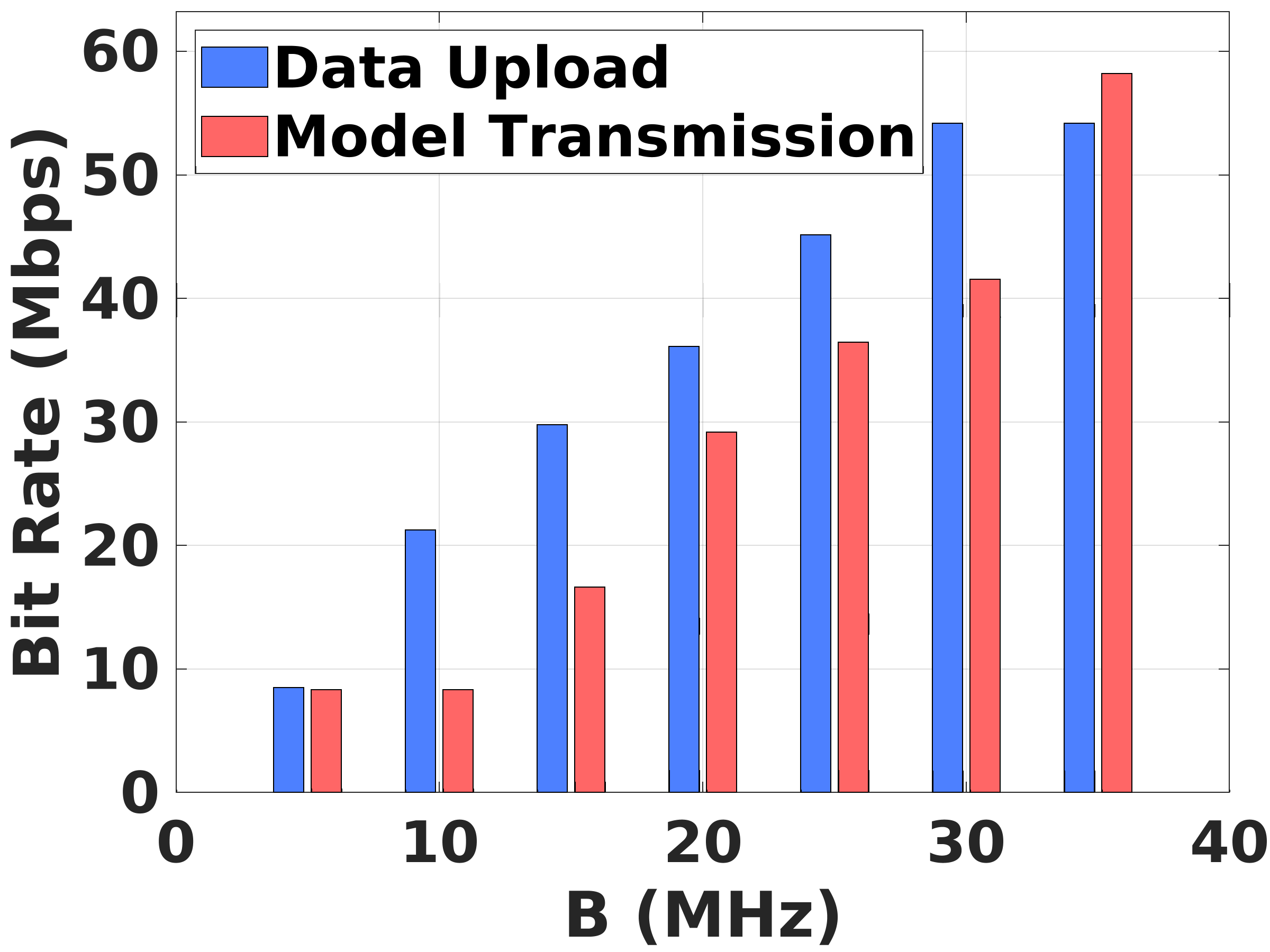}
    \caption{Overhead of uplink data stream \& downlink model stream versus total bandwidth $B$.}
    \label{fig:overheadB}
\end{figure}
Fig.~\ref{fig:bandwidth_map} compares the \( mAP \) of the proposed framework with that of other methods across different bandwidths. The distributed computing framework serves as the baseline. It shows that our method closely matches the exhaustive search results, differing by no more than 0.3\%, thereby validating the proposed algorithm. The CEMA exhibits similar growth at lower bandwidths but performs 2.5\% worse than our method at bandwidths above 20 MHz. In summary, Fig. 2 demonstrates that our algorithm optimally allocates communication resources and outperforms the CEMA.

Fig.~\ref{fig:N_map} compares the \(mAP\) of the proposed framework with that of the CEMA and the distributed computing framework at different frame rates. The distributed computing framework remains the baseline. As the frame rate increases, the \(mAP\) of CEMA continuously decreases, whereas the \(mAP\) of the proposed framework keeps increasing. When $N$ is 80, the $mAP$ of CEMA and the proposed framework are close. When $N$ reaches 200, the proposed framework performs 4.9\% better than CEMA. This is because CEMA follows a relatively fixed upload strategy, leading to insufficient learning as the data volume increases. In contrast, in the proposed framework, when bandwidth is sufficient, $R_V$ increases with \( N \), allowing more data for learning and achieving better model updates. These results indicate that the proposed framework exhibits significant advantages in scenarios where the end device collects a large volume of sensory data.

In Fig.~\ref{fig:overheadB}, we present the trade-off between data upload and model transmission in the proposed framework under various total bandwidths $B$. Under low-bandwidth conditions, increasing bandwidth leads to higher quantization parameters, resulting in a higher proportion of uploaded data. Under medium-bandwidth conditions, the proportion of transmitted model data increases. At high bandwidths, the volume of transmitted model data approaches its limit. Therefore, the proportion of uploaded data continues to rise. 

\section{Conclusion}
This paper has investigated the issue of communication resource allocation in end-edge collaboration system. This system enables small model updates on the end device via real-time communication with a large model deployed on an edge server, which in turn enhances the inference capability of the end model. Given limited bandwidth, we have formulated a joint optimization problem to balance data upload for model training and model transmission for inference enhancement. By solving this optimization problem, we have derived a communication resource allocation algorithm. Simulation results demonstrate that our framework appropriately handles the trade-off between uplink and downlink data transmission, thereby enhancing the inference performance. 

\section*{Appendix A}
\customsubsection{Proof of Theorem 1}
\( L_M(M) \) is the solution to subproblem \eqref{eq:subproblem1}. The solution method follows the approach described in Section IV. The first step involves solving the concave optimization problem for each \( q_j \in \bm{\Omega} \), where \( j \in \{1, 2, 3, \dots, |\bm{\Omega}|\} \). Since any \(g_j(\rho)\) is a concave function, it has a unique maximum value within the closed interval \([0, 1]\), with the corresponding values of \(\rho\) denoted as \(\rho_{\text{best},j}\). The solutions are as follows:
\begin{equation}
0\leq M \leq \frac{T_{total}S_d}{b}(B-\frac{\rho_{\text{best},j}q_jNF}{S_u})\label{situation1}
\end{equation}
\begin{equation}
\frac{T_{total}S_d}{b}(B-\frac{\rho_{\text{best},j}q_jNF}{S_u})\leq M \leq\frac{T_{\text{total}} \cdot B \cdot S_d}{b}\label{situation2}
\end{equation}

\begin{equation}
L_{M,j}(M) =
\begin{cases}
g_j(\rho_{\text{best},j}), & \text{if } \eqref{situation1}. \\
g_j\left(\frac{Su}{NFq_j}(B-\frac{Mb}{T_{total}S_d}) \right), & \text{if } \eqref{situation2}.
\end{cases}
\label{eq:theorem1_LTj}
\end{equation}

It follows that any function \( L_{M,j}(M) \) is continuous and non-decreasing, consisting of a constant function in the first segment and a concave function in the second. Next, we compare all \( L_{M,j}(M) \) functions to derive \( L_M(M)\) and the corresponding values of \( q \). 
\begin{equation}
L_M(M) = \max_{j} L_{M,j}(M),  \label{eq:LM}
\end{equation}

Hence, we have proven that \( L_M(M) \) is a continuous, monotonically non-decreasing piecewise function defined on \( [0, \min(M_{\text{max}}, \frac{T_{\text{total}} \cdot B \cdot S_d}{b})] \), with each segment being either a constant or concave function.

\section*{Appendix B}
\customsubsection{Proof of Theorem 2}
Assume that \((M_{\text{opt}}, mAP^*_{\text{opt}})\) is the optimal solution of Problem \eqref{eq:simplified_objective}. By the definition of the function \( L_M(M) \), for all \( M \in [0, M_{\text{max}}] \), it holds that \( L_M(M) \geq mAP^*_{\text{opt}} \). Suppose, for contradiction, that \( mAP^*_{\text{opt}} \neq L_M(M_{\text{opt}}) \). Then there is \(L_M(M_{\text{opt}})>mAP^*_{\text{opt}}\). Denote the model performance without update as $mAP_{pre}$, there is:
\begin{equation}
h(0, mAP^*) \equiv mAP_{\text{pre}},\quad 
h(M_{\text{max}}, mAP^*) \equiv mAP^*.
 \label{proof2}
\end{equation}
Based on \eqref{proof2} and remark 1, condition \(L_M(M_{\text{opt}})>mAP^*_{\text{opt}} \) leads to a contradiction. Therefore, Theorem 2 holds.

\end{document}